# Dual-Frequency Absorption Spectroscopy in Laser-Cooled Rubidium Atoms: Theoretical Modeling and Experiment

Gour Pati[1,*], Mauricio Pulido[1], Fredrik K. Fatemi[2], Gustavo Acosta[1], and Renu Tripathi[1]

[1]Division of Physics, Engineering, Mathematics & Computer Science, Delaware State University, Dover, Delaware 19901, USA
[2]DEVCOM Army Research Laboratory, Adelphi, Maryland 20783, USA
*Correspondence author: gspati@desu.edu

## ABSTRACT

We demonstrate dual-frequency absorption spectroscopy (DFAS) using laser-cooled $^{87}$Rb and $^{85}$Rb atoms. Doppler-free resonances with high-contrast are produced, which suggest the suitability of using dual-frequency absorption spectroscopy (DFAS) for laser stabilization in a cold-atom-based coherent population trapping (CPT) clock, and for developing a compact, high-performance optical frequency standard using an integrated magneto-optical trap (MOT). We developed a model using density-matrix equations to accurately simulate DFAS in the atomic medium without applying any simplifying approximations. Comprehensive simulations are performed using our multi-level system model to analyze dual-frequency spectra produced in cold atom ensembles under different experimental conditions, including the effect of magnetic field, and two-photon detuning. The simulations accurately yield amplitudes, linewidths, frequency shifts, and lineshapes of DFAS resonances under these experimental conditions. We also demonstrate a simple mechanism for performing CPT spectroscopy by implementing a DFAS laser lock using trapped atoms in the MOT. Additionally, we have extended our model to accurately model the dual-frequency spectrum produced in rubidium cell, which is a medium of practical interest for vapor-cell-based quantum sensing applications.

## I. INTRODUCTION

Laser stabilization is a key requirement in most atomic, molecular and quantum optical experiments. Different spectroscopic techniques have been developed for this purpose [1-10]. The most common ones used in AMO experiments are saturated-absorption spectroscopy (SAS), polarization spectroscopy (PS) and modulation transfer spectroscopy (MTS) [1,2,6-10]. These techniques use a counter-propagating pump-probe geometry to create Doppler-free resonances corresponding to the atomic transitions. Recently, another spectroscopic technique, known as dual-frequency absorption spectroscopy (DFAS) has been developed in the context of laser stabilization for a coherent population trapping (CPT) clock [11,12]. Laser stabilization is important for achieving high long-term clock stability. Typically, a CPT clock uses lasers that are directly modulated instead of external modulation components to keep its architecture compact [13,14]. However, this is associated with the difficulty that the modulated laser cannot be stabilized using the conventional SAS method, particularly when the laser (or carrier) frequency is away from resonance [15]. In this situation, DFAS provides an excellent mechanism for stabilizing the laser [12].

Unlike SAS, DFAS uses counter-propagation of dual-frequency optical fields that satisfy the two-photon resonance condition. Thus, resonances produced by DFAS are influenced by the CPT effect. For the case where the counter-propagating optical fields have orthogonal polarizations, the CPT effects produced by the counter-propagating beams can cancel each other [11]. While SAS exhibits Doppler-free resonances with decreased absorption, cancellation of CPT effects in DFAS results in increased absorption near the line center, which is observed by scanning the carrier frequency of the modulated laser. So far, DFAS has only been studied in Cs and Rb warm vapor media, for improving the long-term stability of a Cs cell CPT clock, and also with motivation towards development of a microcell-based optical frequency standard [11,12,16-18]. Theoretical models have been developed for studying DFAS [17,19]. However, these models are based on a simplified three-level system and give approximate results using calculations that involve perturbative expansion of density-matrix elements in terms of spatial harmonics. Furthermore, the simplified three-level system model cannot be used to accurately model the lineshapes of resonances produced by DFAS, nor can it be used to study the effect of light polarization and magnetic field on the spectral characteristics of these resonances.

In this paper, we develop a multi-level system model to simulate DFAS in cold-atom as well as in warm vapor, and extend experimental DFAS investigation to the cold-atom regime using laser-cooled rubidium atoms in a magneto-optical trap (MOT). Due to the reduced velocity spread in a cold-atom system, our study shows that high-contrast Doppler-free DFAS resonances with near-Lorentzian lineshapes and narrow linewidths (close to the natural-linewidth) are produced using very weak (or low-intensity) excitation. This is promising for laser stabilization as well as for potential development of cold-atom-based optical frequency standards. A detailed numerical simulation of DFAS is also presented using the atomic density-matrix model. In this context, we have developed a new framework for constructing density-matrix equations for the counter-propagating beam geometry. Our multi-level system model includes all relevant energy levels in the rubidium D1 manifold. Unlike the simplified three-level system model, the multi-level system model allows us to study the effect light polarization. Spectral characteristics of DFAS resonances produced by laser-cooled $^{85}$Rb and $^{87}$Rb atoms are studied theoretically and experimentally to show the effect of two-photon detuning and external magnetic field. We demonstrate CPT spectroscopy using our cold-atom system by performing laser stabilization using the DFAS resonance produced by trapped atoms in the MOT. This is relevant to our ongoing effort towards developing a cold-atom-based spin-squeezing-enhanced CPT clock. Additionally, we have extended our multi-level system model to simulate DFAS in warm rubidium vapor. In this case, the simulation accurately predicts the lineshape of dual-frequency spectroscopy produced in a vapor cell.

The rest of this paper is organized as follows. In Section II, we present the theoretical model for DFAS using density-matrix equations. This is described first using a simple three-level Lambda system, and later, using a comprehensive multi-level atomic system. In Section III, we describe our experimental setup used for studying DFAS phenomenon in laser-cooled rubidium atoms. In Section IV, we discuss the results obtained from experiment and theory, drawing comparisons between them. This is followed by conclusions in Section V.

## II.  THEORETICAL MODEL

In this section, we discuss a new method for formulating density-matrix equations for DFAS involving excitation of the atomic medium with counterpropagating optical fields. Simulation is performed by finding numerical solutions to these density-matrix equations under the steady-state condition without using any approximation. We illustrate this by first carrying out DFAS simulations using a simple three-level Lambda system. This is then extended to a multi-level atomic system including all relevant energy levels in the rubidium D1 manifold for accurate modeling of DFAS.

### A.  Three-level Lambda system model

Figure 1 shows a three-level Lambda system with interacting dual-frequency counterpropagating optical fields. This constitutes an idealized system of energy levels in the rubidium D1 manifold where states $|1\rangle$ and $|2\rangle$ can be thought of as the two hyperfine ground-state $F$ levels in $5S_{1/2}$ and state $|3\rangle$ as one of the hyperfine excited-state $F'$ levels in $5P_{1/2}$. Using the complex plane-wave form, the counterpropagating dual-frequency optical fields can be described as

$$\begin{aligned}\vec{E}(z,t) &= \vec{E}^{(+)}(z,t) + \vec{E}^{(-)}(z,t) \\ &= \hat{e}\left[E_1^{(+)}e^{-(\omega_1 t - k_1 z)} + E_2^{(+)}e^{-(\omega_2 t - k_2 z)}\right] \\ &\quad + \hat{e}'\left[E_1^{(-)}e^{-(\omega_1 t + k_1 z)} + E_2^{(-)}e^{-(\omega_2 t + k_2 z)}\right]\end{aligned} \quad (1)$$

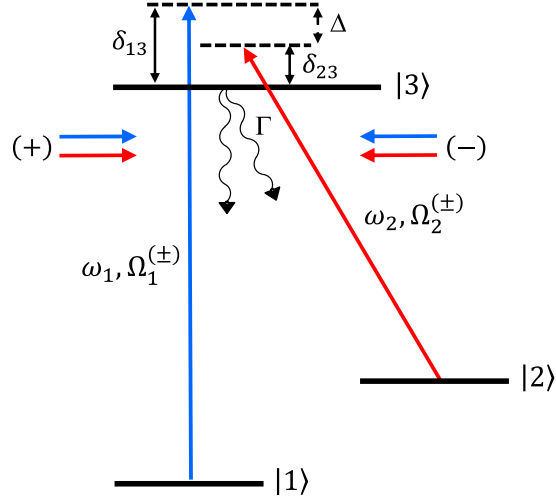

FIG. 1. Energy level diagram showing a three-level Lambda system driven by counter-propagating dual-frequency optical fields.

where the symbol $(\pm)$ corresponds to counterpropagating waves in the $+z$ and $-z$ directions, respectively. $E_1^{(\pm)}$ and $E_2^{(\pm)}$ are the amplitudes of $(\pm)$ optical fields with frequencies $\omega_1$ and $\omega_2$, and propagation constants $k_1$ and $k_2$, respectively. Similarly, $\hat{e}$ and $\hat{e}'$ represent the polarizations of the counterpropagating wave fields. We first model the interactions of counterpropagating $(\pm)$ optical fields separately using the Liouville density-matrix equations [20, 21]. Applying the rotating-wave approximation (RWA), these equations are given as follows:

$$\frac{\partial \rho^{(\pm)}}{\partial t} = -\frac{i}{\hbar}[H^{(\pm)}, \rho^{(\pm)}] - \frac{1}{2}\{\hat{\Lambda}, \rho^{(\pm)}\} + \hat{R}$$

where
$$H^{(\pm)} = \hbar \begin{pmatrix} \delta_{13}^{(\pm)} & 0 & -\frac{\Omega_1^{(\pm)}}{2} \\ 0 & \delta_{23}^{(\pm)} & -\frac{\Omega_2^{(\pm)}}{2} \\ -\frac{\Omega_1^{(\pm)}}{2} & -\frac{\Omega_2^{(\pm)}}{2} & 0 \end{pmatrix}$$

$$\hat{\Lambda} = \begin{pmatrix} \gamma & 0 & 0 \\ 0 & \gamma & 0 \\ 0 & 0 & \gamma+\Gamma \end{pmatrix} \text{ and } \hat{R} = \begin{pmatrix} \frac{\gamma}{2}+\frac{\Gamma}{2}\rho_{33} & 0 & 0 \\ 0 & \frac{\gamma}{2}+\frac{\Gamma}{2}\rho_{33} & 0 \\ 0 & 0 & 0 \end{pmatrix} \qquad (2)$$

Here, $\rho^{(\pm)}$ and $H^{(\pm)}$ represent the density operators and Hamiltonians for $(\pm)$ fields, and $\Omega_i^{(\pm)} = (d_i/\hbar).E_i^{(\pm)}(z)$, $i = 1,2$ corresponds to Rabi frequencies with electric fields $E_i^{(\pm)}(z)$ defined as $E_i^{(\pm)}(z) = E_i^{(\pm)} e^{\pm ik_i z}$ and $d_i$ corresponds to the transition-dipole moment between the ground and the excited state. Doppler shifts due to atomic motion are included by defining the frequency detunings of $(\pm)$ fields as $\delta_{13}^{(\pm)} = \delta_{13} \pm k_1 v_z$ and $\delta_{23}^{(\pm)} = \delta_{23} \pm k_2 v_z$, where $\delta_{13}$ and $\delta_{23}$ correspond to detunings for stationary atoms, as shown in Fig. 2 and $v_z$ is the atomic velocity component along the $+z$ direction. The diagonal relaxation matrix $\hat{\Lambda}$ includes the spontaneous decay rate, $\Gamma$ of the excited state, and the transit relaxation rate, $\gamma$ for each level due to exit of atoms from the laser beam. Matrix $\hat{R}$ describes the repopulation of the ground states due to $\Gamma$ and $\gamma$. The density-matrix equations in eqn. (2) are constructed using the Mathematica-based Atomic Density Matrix (ADM) package [22]. Since the density-matrix equations are coupled equations, they can be combined for the $(\pm)$ fields using the following rules [23]: a) density-matrix equations corresponding to the time derivatives of population terms (i.e. $\frac{\partial \rho_{11}}{\partial t}, \frac{\partial \rho_{22}}{\partial t}$, and $\frac{\partial \rho_{33}}{\partial t}$) and atomic coherence terms (i.e. $\frac{\partial \rho_{12}}{\partial t}$ and $\frac{\partial \rho_{21}}{\partial t}$) in eqn. (2) are combined by adding them to each other for $(\pm)$ fields, and b) density-matrix equations corresponding to the time derivatives of optical coherence terms (i.e. $\frac{\partial \rho_{13}}{\partial t}, \frac{\partial \rho_{31}}{\partial t}, \frac{\partial \rho_{23}}{\partial t}$ and $\frac{\partial \rho_{32}}{\partial t}$) in eqn. (2) are kept as such. Thus, our combined set of density-matrix equations used for simulating DFAS in a three-level Lambda system is as follows:

$$\frac{\partial \rho_{11}}{\partial t} = \frac{\partial \rho_{11}^{(+)}}{\partial t} + \frac{\partial \rho_{11}^{(-)}}{\partial t} = -\left(i\frac{\Omega_1^{(+)}}{2} \rho_{13}^{(+)} + \text{c.c.}\right) - \left(i\frac{\Omega_1^{(-)}}{2} \rho_{13}^{(-)} + \text{c.c.}\right) + \frac{\Gamma}{2}\rho_{33} - \gamma \rho_{11} + \frac{\gamma}{2}$$

$$\frac{\partial \rho_{22}}{\partial t} = \frac{\partial \rho_{22}^{(+)}}{\partial t} + \frac{\partial \rho_{22}^{(-)}}{\partial t} = -\left(i\frac{\Omega_2^{(+)}}{2} \rho_{23}^{(+)} + \text{c.c}\right) - \left(i\frac{\Omega_2^{(-)}}{2} \rho_{23}^{(-)} + \text{c.c}\right) + \frac{\Gamma}{2}\rho_{33} - \gamma \rho_{22} + \frac{\gamma}{2}$$

$$\frac{\partial \rho_{33}}{\partial t} = \frac{\partial \rho_{33}^{(+)}}{\partial t} + \frac{\partial \rho_{33}^{(-)}}{\partial t} = \left(i\frac{\Omega_1^{(+)}}{2} \rho_{13}^{(+)} + \text{c.c.}\right) + \left(i\frac{\Omega_1^{(-)}}{2} \rho_{13}^{(-)} + \text{c.c}\right) + \left(i\frac{\Omega_2^{(+)}}{2} \rho_{23}^{(+)} + \text{c.c.}\right) + \left(i\frac{\Omega_2^{(-)}}{2} \rho_{23}^{(-)} + \text{c.c}\right) - \Gamma\rho_{33} - \gamma$$

$$\frac{\partial \rho_{12}}{\partial t} = \frac{\partial \rho_{12}^{(+)}}{\partial t} + \frac{\partial \rho_{12}^{(-)}}{\partial t} = -i\frac{\Omega_2^{(+)}}{2} \rho_{13}^{(+)} - i\frac{\Omega_2^{(-)}}{2} \rho_{13}^{(-)} + i\frac{\Omega_1^{(+)*}}{2} \rho_{32}^{(+)} + i\frac{\Omega_1^{(-)*}}{2} \rho_{32}^{(-)} - [\gamma + 2i(\delta_{23} - \delta_{13})] \rho_{12}$$

$$\frac{\partial \rho_{21}}{\partial t} = \frac{\partial \rho_{21}^{(+)}}{\partial t} + \frac{\partial \rho_{21}^{(-)}}{\partial t} = \frac{\partial \rho_{12}^*}{\partial t} = i\frac{\Omega_2^{(+)*}}{2} \rho_{31}^{(+)} + i\frac{\Omega_2^{(-)*}}{2} \rho_{31}^{(-)} - i\frac{\Omega_1^{(+)}}{2} \rho_{23}^{(+)} - i\frac{\Omega_1^{(-)}}{2} \rho_{23}^{(-)} - [\gamma - 2i(\delta_{23} - \delta_{13})] \rho_{21}$$

$$\frac{\partial \rho_{13}^{(+)}}{\partial t} = -i\frac{\Omega_1^{(+)}}{2}(\rho_{11} - \rho_{33}) - i\frac{\Omega_2^{(+)}}{2}\rho_{12} - \left(\frac{\Gamma}{2} + \frac{\gamma}{2} + i\delta_{13}^{(+)}\right)\rho_{13}^{(+)}$$

$$\frac{\partial \rho_{13}^{(-)}}{\partial t} = -i\frac{\Omega_1^{(-)}}{2}(\rho_{11} - \rho_{33}) - i\frac{\Omega_2^{(-)}}{2}\rho_{12} - \left(\frac{\Gamma}{2} + \frac{\gamma}{2} + i\delta_{13}^{(-)}\right)\rho_{13}^{(-)}$$

$$\frac{\partial \rho_{31}^{(+)}}{\partial t} = \left[\frac{\partial \rho_{31}^{(+)}}{\partial t}\right]^*; \quad \frac{\partial \rho_{31}^{(-)}}{\partial t} = \left[\frac{\partial \rho_{31}^{(-)}}{\partial t}\right]^*$$

$$\frac{\partial \rho_{23}^{(+)}}{\partial t} = -i\frac{\Omega_2^{(+)}}{2}(\rho_{22} - \rho_{33}) - i\frac{\Omega_1^{(+)}}{2}\rho_{21} - \left(\frac{\Gamma}{2} + \frac{\gamma}{2} + i\delta_{23}^{(+)}\right)\rho_{23}^{(+)}$$

$$\frac{\partial \rho_{23}^{(-)}}{\partial t} = -i\frac{\Omega_2^{(-)}}{2}(\rho_{22} - \rho_{33}) - i\frac{\Omega_1^{(-)}}{2}\rho_{21} - \left(\frac{\Gamma}{2} + \frac{\gamma}{2} + i\delta_{23}^{(-)}\right)\rho_{23}^{(-)}$$

$$\frac{\partial \rho_{32}^{(+)}}{\partial t} = \left[\frac{\partial \rho_{32}^{(+)}}{\partial t}\right]^*; \quad \frac{\partial \rho_{32}^{(-)}}{\partial t} = \left[\frac{\partial \rho_{32}^{(-)}}{\partial t}\right]^* \quad (3)$$

The decay and relaxation rates in the combined equations are modified and reduced to half to keep the total decay rates of the closed system equal to $\Gamma$ and $\gamma$, which keep the atomic population conserved. We solved the system of eqns. (3) under the steady-state condition (i.e. $\frac{\partial \rho}{\partial t} = 0$). This is done by casting the linear equations from eqn. (3) in the matrix-vector form, $MA = B$, where $M$ is a 13×13 coefficient matrix, $A$ is $(13 \times 1)$ column density vector of the form $[\rho_{11}, \rho_{22}, \rho_{33}, \rho_{12}, \rho_{21}, \rho_{13}^{(+)}, \rho_{13}^{(-)}, \rho_{31}^{(+)}, \rho_{31}^{(-)}, \rho_{23}^{(+)}, \rho_{23}^{(-)}, \rho_{32}^{(+)}, \rho_{32}^{(-)}]^T$ and $B$ is $(13 \times 1)$ non-zero constant vector of the form $[-\frac{\gamma}{2}, -\frac{\gamma}{2}, 0, 0, 0, 0, 0, 0, 0, 0, 0, 0, 0]^T$, where $T$ indicates the transpose. Elements of the density vector, $A$, are calculated by numerically inverting the matrix $M$ and multiplying it with the vector $B$. To calculate the light absorption coefficient, we defined the observable, $\alpha = \alpha^{(+)} + \alpha^{(-)} = \frac{3N_a \Gamma \lambda^2}{8\pi} \sum_{i=1}^{2} \left( Im[\rho_{i3}^{(+)}]/|\Omega_i^{(+)}| + Im[\rho_{i3}^{(-)}]/|\Omega_i^{(-)}| \right)$. Here, $N_a$ represents the atom number density and $\lambda$ represents the rubidium D1 transition wavelength. A negative (or positive) value of $\alpha$ will correspond to absorption (or amplification).

Figure 2 shows DFAS resonances calculated using this absorption coefficient, $\alpha$ as a function of laser frequency detuning $\delta$. The two-photon resonant condition is established by choosing $\delta_{13} = \delta_{23} = \delta$ (or difference detuning, $\Delta = \delta_{13} - \delta_{23} = 0$) and by considering stationary atoms in the simulation. We studied the following three cases for calculating DFAS resonances: excitation a) with forward-propagating $(+)$ fields only, b) with counter-propagating $(\pm)$ fields and without applying the CPT cancellation effect, and c) with counter-propagating $(\pm)$ fields and applying the

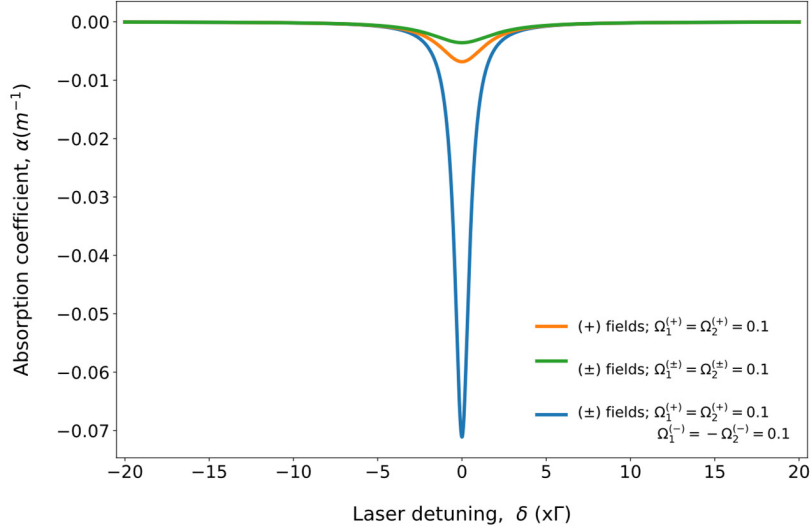

FIG. 2. DFAS resonances calculated for stationary atoms using the three-level Lambda system model for the following three cases: excitation A) with (+) fields only (orange trace), B) with (±) fields and without applying CPT cancellation effect (green trace), and C) with (±) fields and applying CPT cancellation effect (blue trace). Other parameters used in simulation are $\Gamma = 1$, $\gamma = 0.001$ and $\delta_{13} = \delta_{23} = \delta$.

CPT cancellation effect described below. As shown in Fig. 2, in all three cases, inverted resonances (indicating increased absorption) are produced with dips centered around $\delta = 0$ and measured linewidths approximately equal to $\Gamma$. In this simulation, the numerical value of $\Gamma$ is set equal to unity and the magnitudes of Rabi frequencies (i.e. $\left|\Omega_i^{(\pm)}\right|$, $i = 1,2$) are set equal to $\Gamma/10 = 0.1$ (as shown in the legend in Fig. 2). Since the frequencies involved in DFAS are two-photon resonant (i.e. $\Delta = 0$), absorption near the line center (i.e. $\delta = 0$) is strongly influenced by the dark state created by CPT. Away from the line center, the absorption is primarily determined by single-photon absorption associated with the individual frequency components present in the optical fields. Counter-propagation drastically modifies the absorption produced near the line center because dark states created by forward-propagating and counterpropagating optical fields can either add (or cancel) to reduce (or enhance) the absorption effect. Addition or cancellation would very much depend on the polarization states of the optical fields which would typically decide the relative phase (or angle) between the dark states. In actual experiment, cancellation of the dark state (or CPT effect) occurs by choosing orthogonal linear polarization states for the forward-propagating (+) and counter-propagating (−) optical fields. Since polarization states of optical fields cannot be actually simulated in our idealized three-level model, we deliberately modeled this effect by changing the sign of Rabi frequencies for the counter-propagating fields in the following manner: $\Omega_1^{(+)} = \Omega_1^{(-)} = 0.1$ and $\Omega_2^{(+)} = -\Omega_2^{(-)} = 0.1$. Case-C in Fig. 2 shows significantly enhanced absorption near the line center due to perfect cancellation of the dark state which happens by changing the sign of Rabi frequencies in the manner described above. Later on, we demonstrate this effect in the rubidium D1 line using a multi-level system model by choosing orthogonal linear polarization states for (+) and (−) optical fields, similar to how it is done in our actual experiment described in Sec. III.

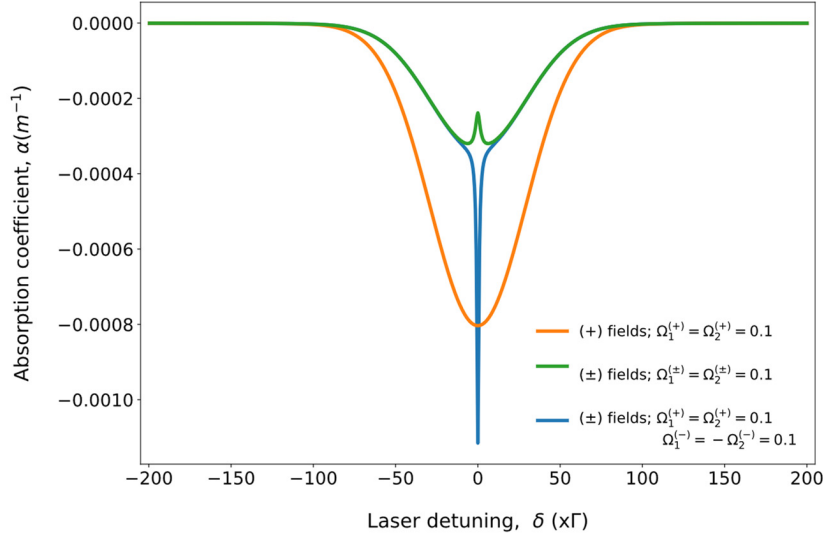

FIG. 3. DFAS resonances calculated by including velocity averaging in the three-level Lambda system model, for following three cases: a) with (+) fields only (orange trace), b) with (±) fields and without CPT cancellation effect (green trace), and c) with (±) fields and CPT cancellation effect (blue trace). Other parameters used in simulation are the same as in Fig. 2.

Next, we extended our three-level Lambda system model to include velocity averaging for modeling DFAS resonances in a Doppler medium. Figure 3 shows DFAS resonances similar to the three cases in Fig. 2, which are obtained by calculating the velocity-averaged absorption coefficient, $\alpha$, as a function of laser detuning, $\delta$. Velocity averaging is performed by introducing velocity-dependent single-photon detuning terms $\delta_{13}^{(\pm)}$ and $\delta_{13}^{(\pm)}$ in Eqns. (3) and calculating a weighted-average of the absorption coefficient, $\alpha(v_z)$ by integrating over a range of velocity groups. The weighting is done using a 1D Maxwell-Boltzmann velocity distribution function described by $(1/\sqrt{2\pi}\, v_p) \exp(-v_z^2/v_p^2)$, where $v_p = \sqrt{2k_B T/m}$ is the most probable velocity corresponding to atomic mass, $m$, and sample temperature, $T$ and $k_B$ is the Boltzmann constant. In the absence of counter-propagation, velocity averaging leads to a Doppler-broadened absorption profile with a width approximately equal to 100Γ, as shown in case-A (orange trace) in Fig. 3. However, in the presence of counter-propagation, the DFAS resonance with a sub-Doppler feature is observed at the center of Doppler-broadened absorption profile. This is due to counter-propagation of optical fields, which is velocity-selective, and a sub-Doppler resonance is created by the resonant zero-velocity group atoms. By choosing the condition for CPT cancellation (i.e.

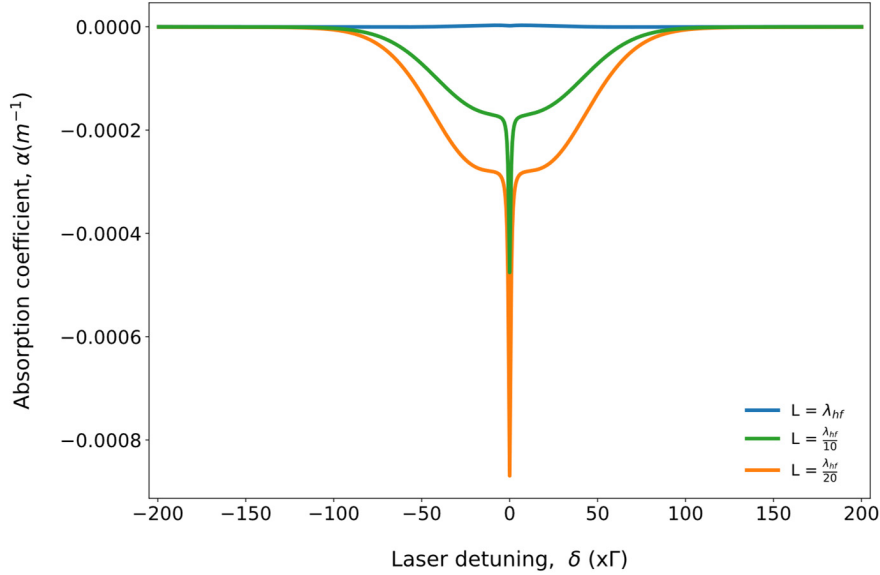

FIG. 4. DFAS resonances calculated using the three-level Lambda system model, showing effect of medium length. Amplitude of the resonance dip is found to be reduced drastically by increasing the medium length, $L$ in comparison to $\lambda_{hf}$ in our simulation. Other parameters used in simulation are the same as in Fig. 2.

$\Omega_1^{(+)} = \Omega_1^{(-)} = 0.1$ and $\Omega_2^{(+)} = -\Omega_2^{(-)} = 0.1$), a pronounced absorption dip is produced as a sub-Doppler feature at the center Doppler-broadened absorption profile. Such a characteristic of DFAS resonance observed using our simple three-level simulation matches well with experimental results reported in a rubidium cell [18]. By choosing the condition $\Omega_1^{(\pm)} = \Omega_2^{(\pm)} = \Omega$, which leads to coherent addition of CPT effect, our simulation in Fig. 3 shows that the sub-Doppler DFAS resonance feature manifests as a peak (instead of a dip) inside the Doppler-broadened absorption profile.

In addition to simulating the Doppler effect, we also extended our simulation to study the effect on the DFAS resonance due to finite length of the medium. In this case, the observable, $\alpha$, is integrated over the length, $L$, of the atomic medium along the $z$-direction, assuming constant atom number density, $N_a$. To ensure the propagation of optical fields to be symmetric, we defined $z$ to be zero at the center of the medium. Fig. 4 shows reduction in the amplitude of the DFAS resonance due to increase in $L$ from $\lambda_{hf}/20$ to $\lambda_{hf}$, where the hyperfine wavelength, $\lambda_{hf} \simeq 9.9\ cm$ for $^{85}$Rb and $\lambda_{hf} \simeq 4.4\ cm$ for $^{87}$Rb. This suggests that cells with short length (i.e. $L \ll$

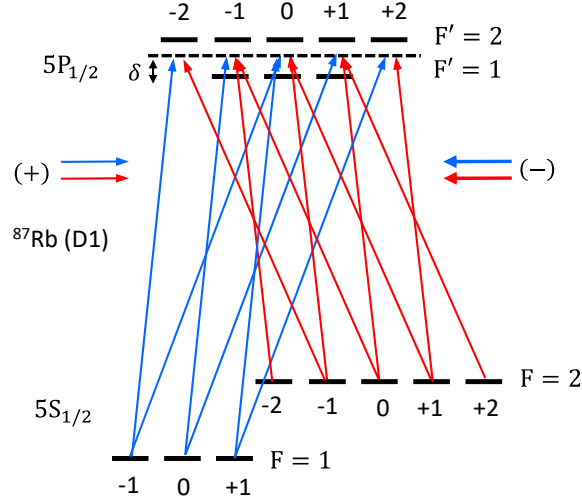

FIG. 5. Energy level diagram showing the multi-level system consisting of 16 energy levels in the D1 manifold of $^{87}$Rb. The diagram shows Lambda subsystems formed by excitation using counter-propagating dual-frequency optical fields. These Lambda subsystems are created by $\sigma^+$ and $\sigma^-$ components of linearly polarized optical fields.

$\lambda_{hf}$) will produce stronger resonance signals. Such an effect has also been observed experimentally using microfabricated cells [17,19]. It occurs due to the fact that counter-propagating dual-frequency optical fields create a spatial oscillation in optical coherence with frequency, $\Delta k = |k_1 - k_2| = 2\pi/\lambda_{hf}$ along the z-direction, and this leads to oscillations in the relative phase between the dark states formed by ($\pm$) fields. The absorption induced by dual-frequency fields is large when $L$ is significantly small compared to the period of this spatial oscillation. The results discussed in Figs. 2-4 illustrate the efficacy of our model in simulating the DFAS phenomenon under different conditions.

**B. Multi-level system model**

The three-level model presented above is approximate. We constructed a comprehensive multi-level system model for investigating the DFAS phenomenon in laser-cooled rubidium atoms. This model includes all relevant energy levels in rubidium atoms corresponding to the D1 manifold. Two separate multi-level models for $^{85}$Rb and $^{87}$Rb atoms were created for comparing our simulation results with actual experiments conducted using $^{85}$Rb and $^{87}$Rb MOTs. The number of energy levels included in the $^{85}$Rb model is 24 and in $^{87}$Rb model is 16. As an example, Fig. 5 illustrates the energy levels of $^{87}$Rb corresponding to the D1 manifold, and shows all possible two-photon transitions (or Lambda systems) created by circular i.e. $\sigma^+(\Delta m_F = 1)$ and $\sigma^-(\Delta m_F = -1)$ polarization components of the linearly polarized optical fields used in DFAS. The Hamiltonians, $H^{(\pm)}$, relaxation matrix, $\hat{\Lambda}$ and repopulation matrix, $\hat{R}$ for the multi-level system are formed explicitly using the ADM Mathematica package [22]. Hamiltonians, $H^{(\pm)}$ are constructed using linearly polarized optical fields with arbitrary angles, and also, considering the Zeeman effect due to the presence of an external magnetic field, $B$. The relaxation matrix, $\hat{\Lambda}$ and repopulation matrix, $\hat{R}$ are formed including the transit and spontaneous decay rates with appropriate branching ratios of $F', m_{F'} \to F, m_F$ which specify the decay to Zeeman sublevels in the $5S_{1/2}$ manifold. A system of linear and time-dependent density-matrix equations for $\rho_{ij}^{(\pm)}$ are created by substituting $H^{(\pm)}$, $\hat{\Lambda}$ and $\hat{R}$ matrices in the Liouville equation described in eqn. (2). This system of equations is combined using the same algorithm that we have developed to construct the combined density-matrix equations for DFAS in a three-level system. The equations after combining are cast in the matrix-vector form: $M A = B$ after applying the steady-state condition (i.e. $\dot{\rho} = 0$). Solutions for the density vector, $A$ are obtained by computing $M^{-1}$ numerically. Although the dimension of $M$ matrix is quite large in this case [i.e. (656 X 656) for $^{85}$Rb and (304 X 304) for $^{87}$Rb], matrix inversion is computed efficiently in Mathematica with high numerical accuracy. We then used the imaginary parts of the optical coherence terms in the density vector, $A$ to define the absorption coefficient, $\alpha$. DFAS resonances are then simulated under different conditions by calculating $\alpha$ as a function the laser detuning, $\delta$. These results are presented later in Sec. IV showing comparisons with our experimental results.

### III. EXPERIMENTAL DESCRIPTION

Figure 6(a) shows the schematic diagram of our experimental system for studying DFAS in laser-cooled rubidium atoms. Laser cooling is performed using a double MOT system (doubleMOT, Infleqtion). In this double MOT system, a 2D MOT is used as a source for feeding atoms into the 3D MOT cell, which is located above the 3D MOT cell. Atoms from the 2D MOT cell travel through a silicon pinhole disc into the 3D MOT cell by pushing them using a vertically-directed resonant beam. The layout of the double MOT laser paths is shown with arrows indicating six-beams that are used for laser cooling in the 3D MOT and four additional beams in the 2D MOT. These beams are derived from two external-cavity diode lasers [not shown in Fig. 6(a)] to produce cooling and repump beams that are combined and delivered to both the 2D and 3D MOTs using optical fibers. The cooling and repump lasers are stabilized using the SAS technique in the D2 transition (i.e. $\lambda \simeq 780$ nm) in rubidium cells. The cooling laser is detuned by approximately 15 $MHz$ below the cycling transition (i.e. $F = 3 \rightarrow F' = 4$ for $^{85}$Rb and $F = 2 \rightarrow F' = 3$ for $^{87}$Rb). By tuning the lasers to appropriate cooling and repump transitions, we can easily switch between a $^{85}$Rb MOT and a $^{87}$Rb MOT. For the DFAS experiment, a separate external-cavity diode laser (DLC DL Pro, Toptica, linewidth $< 500\ kHz$) tuned to the rubidium D1 transition (i.e. $\lambda \simeq$ 795 nm) is used. Light from this laser is sent through a fiber-coupled electro-optic modulator (EOM, NIR-MX800-LN-10-00-P-P-FA-FA, Exail). The EOM is driven by a microwave generator at half the rubidium ground-state hyperfine frequency (i.e. $\nu_{hf/2} \simeq 1.5178\ GHz$ for $^{85}$Rb and $\simeq$ 3.4173 GHz for $^{87}$Rb) so that the first-order optical sidebands generated by the EOM are two-photon resonant with their frequency difference matched to $\nu_{hf}$. A bias controller (MBC-DG-LAB-B2, Exail) is used to actively suppress the carrier by approximately 15 $dB$ using a servo. The EOM used in our case is an intensity modulator. Higher order sidebands that are produced within the 10 $GHz$ EOM bandwidth are also actively suppressed by the bias controller. This creates a near-perfect dual-frequency field for doing CPT or DFAS experiments using only the first-order sidebands. Light after the EOM is directed through an acousto-optic modulator (AOM) for switching and generating optical pulses. The AOM is driven by an RF source at $80\ MHz$ that allows the dual-frequency optical fields to be switched-on only during the experimental cycle.

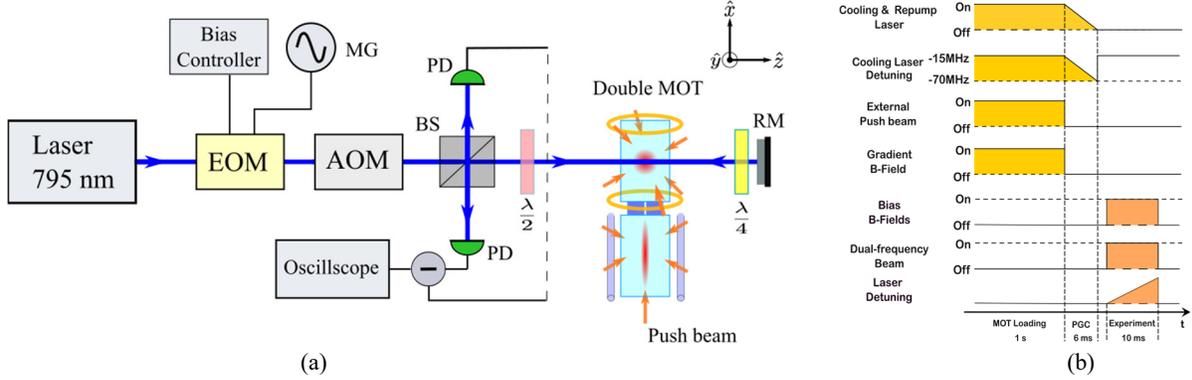

FIG. 6. (a) Layout of the DFAS experimental setup developed using a rubidium double MOT system. The counter-propagating dual-frequency optical fields overlap in the 3D MOT. MG: microwave generator, RM: retroreflecting mirror, BS: beamsplitter, PD: photodetector, $\frac{\lambda}{2}$: half-waveplate, and $\frac{\lambda}{4}$: quarter-waveplate. (b) Timing diagram illustrating the sequences for MOT loading, PGC and the experimental cycle for DFAS.

In our case, DFAS is performed by sending the dual-frequency fields in counter-propagation through the atom cloud produced inside the 3D MOT cell. The counter-propagating beam is generated using the retroreflecting mirror shown in Fig. 6(a). A half-wave ($\lambda/2$) plate is used in the forward beam path to set the polarization state to vertical. A quarter-wave ($\lambda/4$) plate is placed in the return beam path to create either a parallel (i.e. vertical, V) or perpendicular (i.e. horizontal, H) polarization state for the counter-propagating beam, by rotating the $\lambda/4$ waveplate. A beam splitter with 70:30 splitting ratio is used at the input. Light outputs from this beam splitter are directed to two photodetectors whose outputs are subtracted to reduce background and intensity noise from the DFAS signal. The experimental run starts by MOT loading for one second followed by $6\ ms$ of polarization gradient cooling (PGC) to attain a lower temperature of the atom cloud below the Doppler limit [24]. Figure 6(b) shows the timing diagram for our experimental run. PGC is performed using MOT beams by simultaneously ramping down the detuning of the cooling laser and the optical power in both cooling and repump lasers. The optimal PGC temperature was measured to be approximately $40\ \mu K$ for both $^{85}$Rb and $^{87}$Rb atom clouds. The DFAS experiment is conducted using the expanding and free-falling atom cloud formed after the PGC. A time delay of $1\ ms$ is introduced after PGC to account for the delay in shutting off the laser-cooling beams using mechanical shutters for high-extinction. The expansion of atom cloud after PGC only allows a time window of approximately $10\ ms$ for acquiring the DFAS signal. During this period, the dual-frequency beam is turned on, as indicated in Fig. 6(b), and the resonance signals produced by exciting the atom cloud are acquired by scanning the laser frequency. Bias coils [not shown in Fig. 6(a)] were used to null the ambient magnetic field present in the atom cloud environment, and also to apply a magnetic field, $B$ along the direction of the dual-frequency beam to study its effect on DFAS. The experimental setup in Fig. 6(a) is also used to demonstrate CPT spectroscopy by locking the laser using the DFAS resonance produced by trapped atoms during the MOT loading cycle. CPT spectroscopy is performed in a similar manner as DFAS, except by sweeping the microwave frequency during the $10\ ms$ experimental time window.

## IV. RESULTS AND DISCUSSIONS

Figure 7(a) shows Doppler-free DFAS resonances produced by the atom cloud in a $^{87}$Rb MOT using an optical power of approximately 15 $\mu W$ (or excitation intensity, $I_{ex} \simeq 0.1 \; mW/cm^2$) in the dual-frequency beam. The ambient magnetic field present in the atom cloud environment is compensated using two bias coils. Two high-contrast absorption dips are observed in the spectrum by scanning the laser frequency over a wide range. As mentioned in Sec. III, the frequency difference between the first-order EOM generated sidebands is kept matched to the ground-state $^{87}$Rb hyperfine frequency (i.e. $\nu_{hf} \simeq 6.8347 \; GHz$) to create the two-photon resonance condition for DFAS. The resonance dip on the left in Fig. 7(a) is produced when the sidebands during the laser scan become resonant with the lower hyperfine excited state, $|F' = 1\rangle$, and similarly, the absorption dip on the right occurs when the sidebands become resonant with the upper hyperfine excited state, $|F' = 2\rangle$ in $^{87}$Rb D1 line. The frequency separation between these two resonance dips corresponds to the excited-state hyperfine splitting ($\simeq 816 \; MHz$), as indicated in Fig. 7(a). Dual-frequency absorption spectra shown in Fig. 7(a) were acquired by choosing parallel (i.e. VV) as well as orthogonal (or crossed, i.e. VH) linear polarization states for the counter-propagating fields (where V: Vertical and H: Horizontal). In both bases, resonance dips are observed to be prominent, although the amplitudes of resonance dips produced by VH polarization are higher. Using our multi-level system simulation model, we verified that this enhancement most likely occurs due to better cancellation of dark states (or CPT effect) for VH polarization than VV polarization. Due to the reduced velocity spread in laser-cooled atoms, the resonances are observed to have high contrast and near Lorentzian lineshapes without a Doppler-broadened background. In Fig. 7(a), we also show the SAS signal obtained using a reference rubidium cell (containing $^{85}$Rb and $^{87}$Rb mixture). The SAS signal is obtained simultaneously during the laser scan by picking a small portion of the unmodulated laser beam before the EOM and sending it through the reference cell in counter-propagation. The SAS signal indicates the laser position when DFAS resonances are produced using resonant sidebands in laser-cooled $^{87}$Rb atoms. This SAS plot shows that the laser is not actually resonant when DFAS resonances are produced. Thus, DFAS resonances can be used to stabilize the laser frequency while it is off-resonant. Later, we will describe CPT spectroscopy that we performed in our double MOT system by locking the laser using the DFAS resonance produced by trapped atoms in the 3D MOT. This approach is currently being used in laser stabilization for developing a cold-atom-based spin-squeezing-enhanced CPT clock [25].

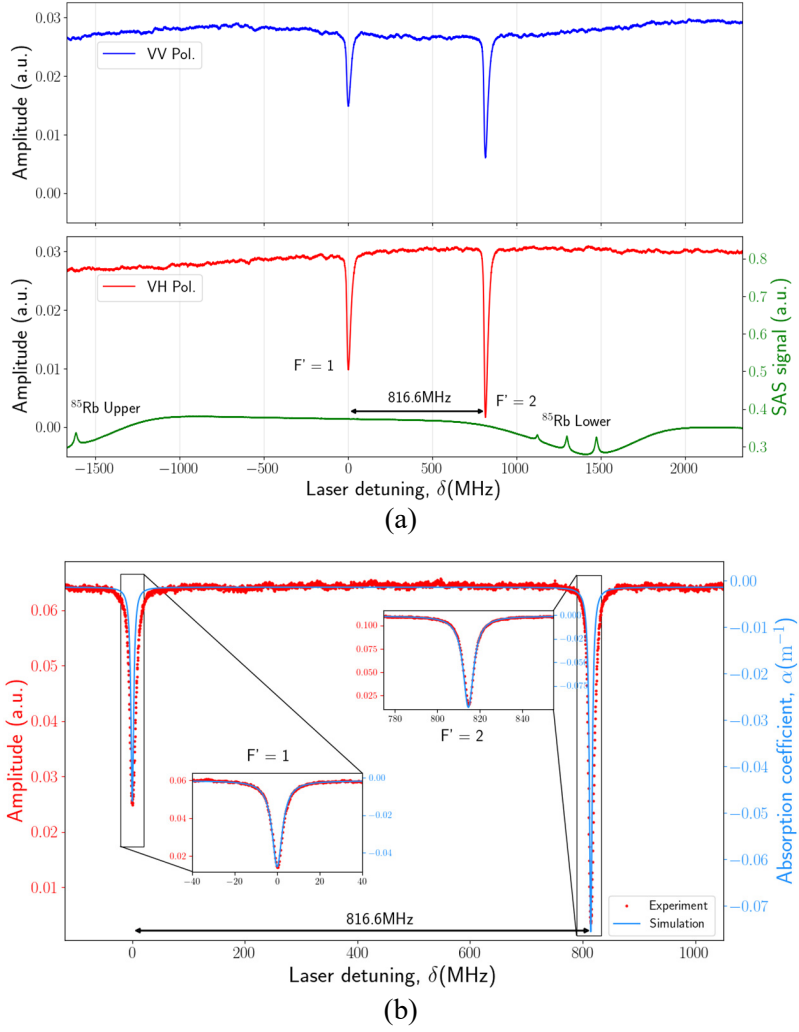

FIG. 7. (a) Dual-frequency absorption spectrum acquired using the $^{87}$Rb atom cloud. Resonance amplitudes are observed to be higher by using counter-propagating fields with crossed polarization states, VH. The green trace shows SAS signal acquired from a reference rubidium cell. (b) Comparison between the dual-frequency absorption spectrum calculated using the multi-level system model and acquired experimentally using a relatively narrow laser scan. Linewidths of resonances observed in this spectrum are found to match closely with the natural linewidth, Γ ($\simeq 5.8\ MHz$) of $^{87}$Rb D1 transition.

Next, we measured the linewidths of DFAS resonances produced in the $^{87}$Rb atom cloud. The linewidths of resonances observed in Fig. 7(a) are artificially broadened due to the wide laser scan. For measuring the true linewidths, we acquired the spectrum using a relatively narrow laser scan of approximately 1.2 $GHz$ which is shown in Fig. 7(b). To compare with theory, we also calculated the resonance spectrum using our muti-level system model with parameters such as Rabi frequencies and polarization states of optical fields same as in the experiment (described in Sec. IIB). The model includes all 16 relevant Zeeman sublevels in the $^{87}$Rb D1 manifold. The simulation shows spectral amplitudes that agree very well with the experiment. In practice, the expansion of the atom cloud during the laser scan causes a small reduction in atomic density. We found that this reduction does not significantly affect the spectral amplitude of $F' = 2$ resonance in Figs. 7(a,b). Linewidths of resonances were accurately measured by fitting the simulated spectrum using Lorentzian functions. They are found to match closely with the natural linewidth, $\Gamma$ ($\simeq 5.8\ MHz$) of $^{87}$Rb D1 transition. To avoid artificial broadening of linewidths in experiment, we acquired individual resonances using significantly narrow laser scans, which are shown in respective insets in Fig. 7(b). The linewidths of these resonances matched very closely with the simulation. Thus, the linewidths of DFAS resonances produced under this weak excitation condition (i.e. $I_{ex} \ll I_{sat}$), are limited by the natural linewidth, $\Gamma$. This may be considered as a drawback of DFAS in comparison to rubidium two-photon spectroscopy which has been found to produce sub-natural linewidth resonances for optical frequency standard development [26,27]. However, the advantage of DFAS could be its relative simplicity in implementation, particularly for optical frequency standard development using a compact and integrated MOT [28].

Figure 8 shows similar spectral measurements performed using the $^{85}$Rb atom cloud. In this case, the frequency difference between the first-order sidebands generated by the EOM, is matched to the $^{85}$Rb ground-state hyperfine frequency (i.e. $\nu_{hf} \simeq 3.0357\ GHz$) to create the two-photon resonance condition for DFAS. A total optical power of approximately 15 $\mu W$ is used in dual-frequency excitation, and the counter-propagating optical fields are cross polarized. Two high-contrast resonance dips corresponding to resonant two-photon transitions with $^{85}$Rb hyperfine excited states, $|F' = 2\rangle$ and $|F' = 3\rangle$ in $^{85}$Rb D1 manifold are observed by scanning the laser frequency. The frequency separation between these two resonance dips corresponds to the excited-state hyperfine splitting which is approximately 362 $MHz$. We also calculated this spectrum using our multi-level system model by including all 24 Zeeman sublevels in the $^{85}$Rb D1 manifold. This is plotted in Fig. 9 for comparison with the experiment. Amplitudes of the resonance dips matched very well with each other. Similar to Fig. 7(b), the insets in Fig. 8 show individual resonances acquired using a narrow laser frequency scan for accurate linewidth measurement. By fitting the resonance lineshapes with Lorentzian functions, linewidths of resonances were measured to be $5.811\ MHz$ for $|F' = 2\rangle$ and $5.876\ MHz$ for $|F' = 3\rangle$, respectively and they are found to match closely with the natural linewidth, $\Gamma$ ($\simeq 5.8\ MHz$) of $^{85}$Rb.

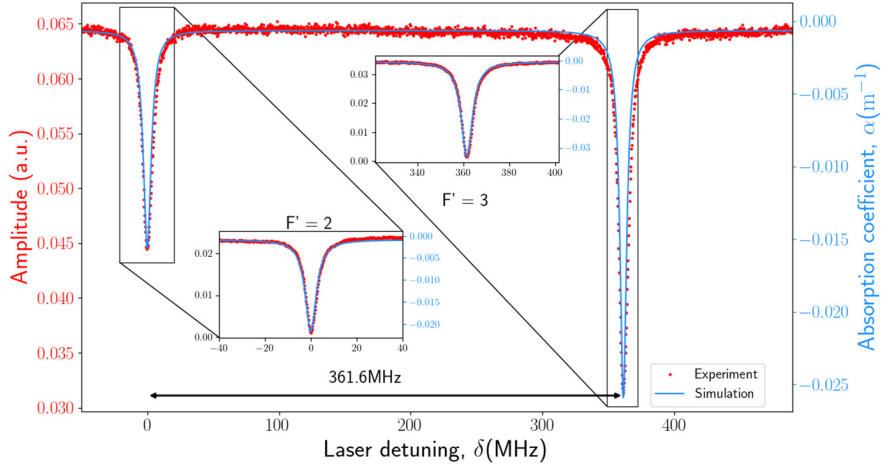

FIG. 8. Comparison between the dual-frequency absorption spectrum calculated using the multi-level system model and acquired experimentally using the $^{85}$Rb atom cloud. The counter-propagating optical fields are cross polarized. The insets show individual resonances acquired using narrow laser frequency scan for accurate linewidth measurement.

Next, we describe the effect of two-photon detuning on DFAS resonances. Figures 9(a,b) contain a sequence of three-dimensional plots showing the individual $F' = 1$ and $F' = 2$ resonances produced by the $^{87}$Rb atom cloud, corresponding to four different values of two-photon (or difference) detuning, $\Delta$ (i.e. $\Delta = 0, 4, 10,$ and $20\ MHz$, respectively). The resonance amplitude is maximum when $\Delta = 0$ and reduces when $\Delta$ increases. For large $\Delta$ values compared to the resonance linewidth, $\Delta\nu$ ($\simeq 5.8\ MHz$), resonance amplitudes are reduced considerably. This shows that enhanced absorption in DFAS resonances is actually produced by the two-photon effect. We simulated this case using our multi-level system model. The simulated resonances are plotted in Figs. 9(a,b) with the experimental results and show strong agreement. Along with reduction in resonance amplitude, the two-photon detuning creates a noticeable frequency shift of the resonance in Figs. 9(a,b) for large $\Delta$ values. We used our simulation to understand the origin of this shift better. It is found that when $\Delta$ becomes large, the resonance is primarily produced by two frequency-shifted single-frequency resonances. For example, in case of $F' = 1$ resonance, the single-frequency transitions $|F = 1\rangle \rightarrow |F' = 1\rangle$ and $|F = 2\rangle \rightarrow |F' = 1\rangle$ create two frequency-shifted single-frequency resonances centered respectively at $\pm \Delta/2$ away from the center of the dual-frequency resonance (i.e. $\delta = 0$). These single-frequency resonances are produced by CPT between the inter-Zeeman sublevels in the $F$ state, and subsequently the absorption induced by CPT cancellation due to counter-propagation of the optical field. The resultant shift of $F' = 1$ resonance in $\Delta \neq 0$ case, is very much dependent on the relative amplitudes of these two underlying single-frequency resonances. Using our simulation, we measured the frequency shifts of $F' = 1$ and $F' = 2$ resonances as a function of $\Delta$ which are shown in Fig. 9(c). By doing a least-squares fitting, we found that the frequency shift of the $F' = 1$ resonance is approximately quadratic, and the frequency shift of $F' = 2$ resonance is approximately linear with respect to $\Delta$. The overall frequency shift of $F' = 2$ resonance was found to be smaller compared to $F' = 1$ resonance, which is possibly due to the fact that excitations to $F' = 2$ produce single-frequency resonances with better matching amplitudes than $F' = 1$. For $\Delta \gg \Delta\nu$, the lineshapes of $F' = 1$ and $F' = 2$ resonances are only governed by single-frequency transitions, which become asymmetric and broader, as seen in Figs. 9(a,b). Similar two-photon detuning effects were also observed on DFAS resonances produced by the $^{85}$Rb atom cloud. Frequency shift and lineshape asymmetry induced by two-photon detuning are important considerations for applications involving laser stabilization and optical frequency referencing.

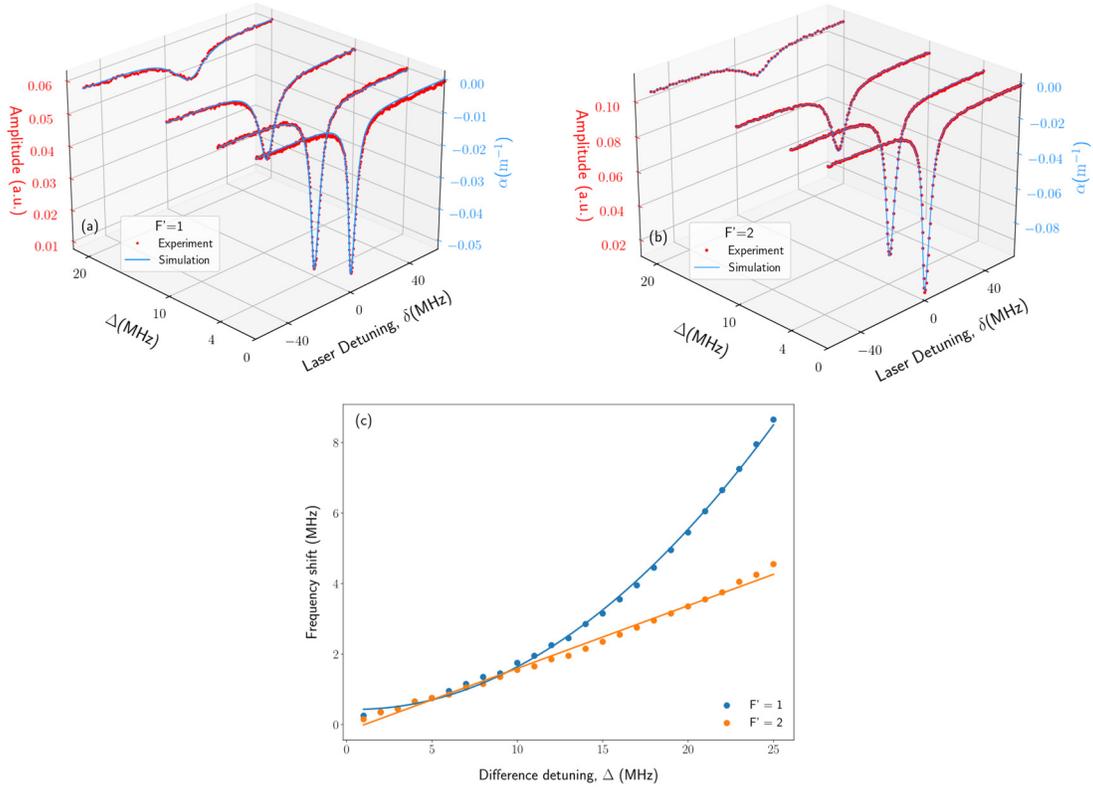

FIG. 9(a,b) Plots showing $F' = 1$ and $F' = 2$ resonances produced by $^{87}$Rb atom cloud, for different values of two-photon (or difference) detuning, $\Delta$. The counter-propagating optical fields are cross polarized. The resonances are accompanied by amplitude reduction and frequency shifts. (c) Frequency shifts of $F' = 1$ and $F' = 2$ resonances plotted as a function of $\Delta$. Solid lines are linear and quadratic fits to the calculated shifts shown using dots.

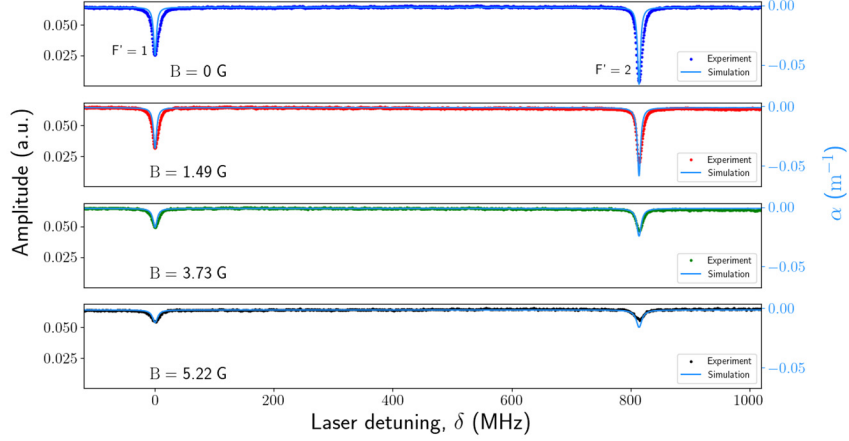

FIG. 10. Plots showing $F' = 1$ and $F' = 2$ resonances produced by $^{87}$Rb atom cloud, for increasing magnetic field strength, $B$. The counter-propagating optical fields are cross polarized. With increasing $B$ field, the resonances show amplitude reduction.

Figure 10 shows the effect of magnetic field on the dual-frequency absorption spectrum of $^{87}$Rb. Using a bias coil, a longitudinal magnetic field up to $5.22 G$ is applied along the propagation direction of the dual-frequency beam [i.e. along the $z-$direction indicated in Fig. 6(a)]. This field is applied only during the experimental cycle of $10\ ms$ as shown in the timing diagram in Fig. 6(b). Amplitudes of both $F' = 1$ and $F' = 2$ resonances are reduced when the magnetic field strength, $B$ is increased. This is because the magnetic field $B$ causes opposite frequency shifts between the hyperfine Zeeman sublevels, and effectively creates a two-photon detuning, the case we discussed earlier in Figs. 9(a,b). Our multi-level system simulation in Fig. 10 predicts similar reduction in resonance amplitudes with increase in $B$. Due to the wide laser scan, the linewidths of $F' = 1$ and $F' = 2$ resonances observed in the experiment are broad in comparison to the simulated resonances. Amplitudes of $F' = 1$ and $F' = 2$ resonances at high $B$ field are produced mainly due to single-frequency transitions, as discussed earlier. Magnetic fields should also cause frequency shifts and lineshape asymmetries similar to the two-photon detuning case. Due to the wide laser scan, these effects cannot be clearly observed. However, they can be more thoroughly and accurately investigated using our multi-level system simulation model.

CPT provides an important tool for cold atom based clock development [29,30]. We performed CPT spectroscopy using DFAS laser lock in our experimental setup shown in Fig. 6(a). This is done by scanning the EOM frequency around $\nu_{hf}$ (which is equivalent to changing $\Delta$ around $\Delta = 0$) during the experimental cycle of $10\ ms$. The subplots in Fig. 11 show CPT spectra obtained respectively from $^{85}$Rb and $^{87}$Rb atom clouds. These are conveniently acquired by locking the laser to the peak of DFAS resonance produced by the trapped atoms that are captured during the MOT loading cycle. CPT spectra shown in Fig. 11 are produced by exciting the atom cloud with dual-frequency beam. During this excitation, the DFAS laser lock is kept engaged by enabling 'integrator hold' of the proportional-integral (PI) servo with a TTL trigger. The laser is locked to $F' = 3$ resonance peak in case of $^{85}$Rb and $F' = 2$ resonance peak in case of $^{87}$Rb. Five CPT resonances are observed in the $^{85}$Rb spectrum, shown in Figs. 11(a) and three CPT resonances are observed in the $^{87}$Rb spectrum, shown in Fig. 12(b) by using $VV$ polarization states for the counter-propagating CPT beams. The resonances are separated by applying the magnetic field, $B$ ($\simeq 380\ mG$) in the longitudinal direction (i.e. $z-$direction). They are formed by multiple dark states which are created by superpositions between the hyperfine magnetic sublevels. The superposition states responsible for forming particular CPT resonances are labeled in Figs. 11(a,b). A total optical power of $15\ \mu W$ is used in doing this CPT spectroscopy. Figures 11(c,d) show that only the CPT end resonances are strongly formed when the beam polarization is chosen to be elliptical. This is a preliminary demonstration which illustrates the convenience of implementing the DFAS laser lock in the cold atom system itself while performing a CPT experiment. The main advantage of using this approach is that the CPT beam in this case does not contain any non-resonant frequency component that can cause frequency error due to light shift in CPT clock [31]. We are currently using this simple laser stabilization method for developing a spin-squeezing-enhanced CPT clock [25]. CPT resonance linewidths in Fig. 11 are broadened due the large $\Delta-$scan rate. In a CPT clock, only the central resonance in the $^{87}$Rb spectrum is relevant. True linewidth of the central CPT resonance can be measured by modifying the experimental run to perform frequent and repeated MOT loadings over a short time period (typically, $50-100\ ms$), followed by discrete stepping of $\Delta$ value around $\Delta=0$. Similarly, contrast in the central CPT resonance can be enhanced using counter-propagating CPT beams with opposite $(\sigma^+-\sigma^-)$ circular polarizations [15,32]. Noise associated with the spectral background can be reduced by improving the design of our DFAS laser lock servo.

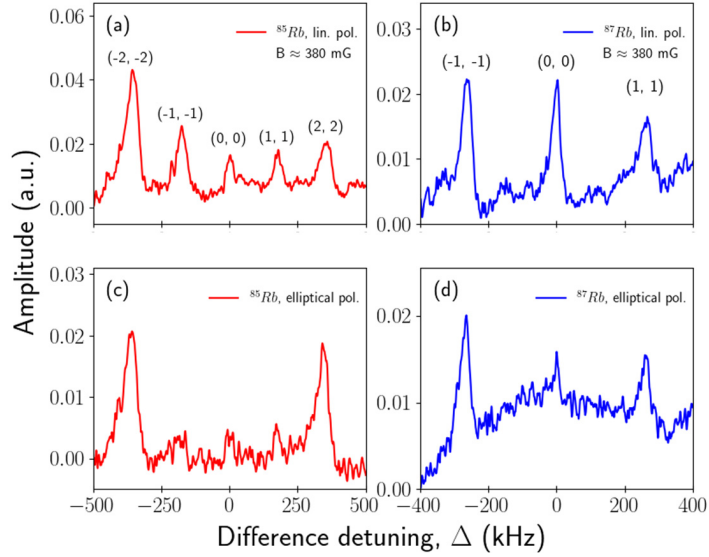

FIG. 11. Plots showing CPT spectra acquired from (a,c) $^{85}$Rb atom cloud, and (b,d) $^{87}$Rb atom cloud by implementing DFAS laser lock with trapped atoms in the MOT. The spectrum consists of magnetically-insensitive resonance at the center and multiple Zeeman CPT resonances. Polarization of CPT beam is chosen to be linear in (a,b) and elliptical in (c,d).

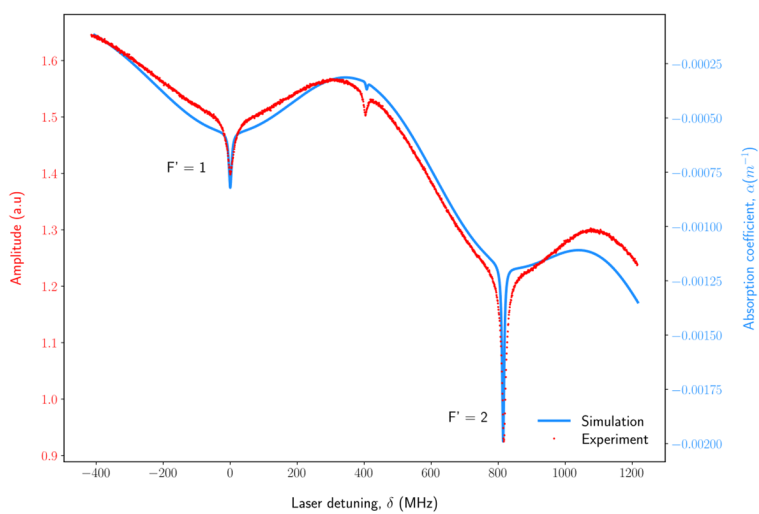

FIG. 12. Dual-frequency spectrum produced in $^{87}$Rb D1 line excitation in a vapor cell. Multi-level system model with velocity-averaging is used to calculate the spectrum to show it as a match with the experiment.

To test the efficacy of our multi-level system model to simulate DFAS in a Doppler medium, we also performed experiments in a rubidium vapor cell. Figure 12 shows the theoretically calculated spectrum along with the experimentally acquired spectrum obtained by performing dual-frequency excitation in a small rubidium cell of length, $1.5\ cm$ and diameter, $2.5\ cm$. The cell is enclosed inside a magnetic-shield cannister to create near-zero magnetic field condition. The spectral measurement shown in Fig. 12 is performed by heating the cell to approximately $70^0C$ and using an optical power of approximately $300\ \mu W$ in the dual-frequency beam. Sub-Doppler resonance dips within the broad Doppler backgrounds are observed using $^{87}$Rb D1 line excitation. Similar to the case of laser-cooled atoms, two resonance dips corresponding to $F'=1$ and $F' = 2$ states are observed by scanning the laser detuning, $\delta$ over a large frequency range. However, the lineshapes of these resonance signals are significantly modified due to the Doppler effect. Our multi-level system simulation predicts this spectrum quite well. The simulation is performed including all sixteen Zeeman sublevels in $^{87}$Rb D1 line and applying velocity-averaging using the 1D Maxwell-Boltzmann velocity distribution. In Fig. 12, we observed an increase in absorption for laser detuning, with $\delta$ exceeding $1\ GHz$. This is due to absorption produced by $^{85}$Rb atoms in the cell. To create this match with experiment, we also included $^{85}$Rb atoms in our simulation. The absorption spectrum produced by $^{85}$Rb atoms is calculated using a detuned excitation by considering the energy (or frequency) difference between $^{87}$Rb and $^{85}$Rb ground states, and the spectrum is then added to the $^{87}$Rb spectrum by multiplying each one of them with respective weighting factors corresponding to their natural abundances. In this case, the calculated spectrum in Fig. 12 showed a similar increase in absorption for $\delta > 1\ GHz$. An additional crossover resonance is formed by non-zero velocity group atoms, exactly midway between the $F' = 1$ and $F' = 2$ resonances. Our multi-level system model can thus accurately simulate DFAS signal in rubidium cell, and can also be extended to short-length cells (i.e. length, $L \ll \frac{\lambda_{hf}}{2} \simeq 2.2\ cm$) including microcells which are relevant to miniaturized quantum sensor development. The model can also be readily extended to study other spectroscopy schemes which involve a counter-propagating geometry similar to DFAS.

## V. CONCLUSIONS

In conclusion, we have investigated experimentally and theoretically DFAS using laser-cooled rubidium atoms. Our study shows that unlike a vapor medium, resonances produced using dual-frequency excitation in cold atoms exhibit high contrast and narrow linewidth without Doppler background. We developed a new theoretical framework to accurately simulate DFAS in cold atoms. A simple three-level system model is used first to explain DFAS. We then performed simulations using the multi-level system model to analyze spectral characteristics of DFAS resonances produced in cold atoms under different experimental conditions. The model is also extended to simulate DFAS in a vapor medium. The simulations show good agreement with experiment and accurately yield amplitudes, linewidths, frequency shifts, and lineshapes of DFAS resonances under various experimental conditions. Our studies demonstrate the viability of DFAS in a cold-atom system for improving performance in laser stabilization, and potentially developing high-performance optical frequency standards. We also demonstrated a simple mechanism for CPT spectroscopy by implementing laser stabilization with DFAS in the cold-atom system. This mechanism is suitable for a cold-atom-based CPT clock.

ACKNOWLEDGMENTS

This work is supported by the Center of Excellence on Advanced Quantum Sensing under the DoD grant W911NF2020276.